\documentclass[prd,twocolumn,reprint,english,aps,preprintnumbers,superscriptaddress,nofootinbib,floatfix]{revtex4-1}

\usepackage{color}
\usepackage{xcolor}
\usepackage{graphicx}
\usepackage{amsfonts}
\usepackage{amssymb,amsmath}
\usepackage{babel}
\usepackage{multirow}
\usepackage{hyperref}
\usepackage{float}
\usepackage{ulem}

\makeatletter

\@ifundefined{definecolor}
{\usepackage{color}}{}

\newcommand{\ba}{\begin{eqnarray}}
\newcommand{\ea}{\end{eqnarray}}
\newcommand{\bef}{\begin{figure}}
	\newcommand{\eef}{\end{figure}}

\usepackage{catchfile}
\newcommand{\getenv}[2][]{%
  \CatchFileEdef{\temp}{"|kpsewhich --var-value #2"}{}%
  \if\relax\detokenize{#1}\relax\temp\else\let#1\temp\fi}
\getenv[\USER]{USER}

\newcommand{\GeV}{\ \text{GeV}}

\newcommand{\be}{\begin{equation}}
\newcommand{\ee}{\end{equation}}
\definecolor{vierde}{rgb}{0.0, 0.5, 0.0}

\newcommand{\dnef}{\Delta N_\text{eff}}

\begin{document}
	
\title{Cosmic Imprints of XENON1T Axions}

\author{Fernando~Arias-Arag\'on}
\email{fernando.arias@uam.es}
\affiliation{Instituto de F\'isica Te\'orica UAM/CSIC, Calle Nicol\'as Cabrera 13-15, Cantoblanco E-28049 Madrid, Spain}
\affiliation{Departamento  de  F\'{\i}sica Te\'{o}rica,  Universidad  Aut\'{o}noma  de  Madrid, Cantoblanco  E-28049  Madrid,  Spain}

\author{Francesco~D'Eramo}
\email{francesco.deramo@pd.infn.it}
\affiliation{Dipartimento di Fisica ed Astronomia, Universit\`a di Padova, Via Marzolo 8, 35131 Padova, Italy}
\affiliation{INFN, Sezione di Padova, Via Marzolo 8, 35131 Padova, Italy}

\author{Ricardo~Z. Ferreira}
\email{ricardo.zambujal@su.se}
\affiliation{Nordita, KTH Royal Institute of Technology and Stockholm University, Roslagstullsbacken 23, SE-106 91 Stockholm, Sweden}

\author{Luca~Merlo}
\email{luca.merlo@uam.es}
\affiliation{Instituto de F\'isica Te\'orica UAM/CSIC, Calle Nicol\'as Cabrera 13-15, Cantoblanco E-28049 Madrid, Spain}
\affiliation{Departamento  de  F\'{\i}sica Te\'{o}rica,  Universidad  Aut\'{o}noma  de  Madrid, Cantoblanco  E-28049  Madrid,  Spain}

\author{Alessio~Notari}
\email{notari@fqa.ub.edu}
\affiliation{Departament de F\'isica Qu\`antica i Astrofis\'ica \& Institut de Ci\`encies del Cosmos (ICCUB), Universitat de Barcelona, Mart\'i i Franqu\`es 1, 08028 Barcelona, Spain}

\preprint{\vbox{\hbox{FTUAM-20-12, IFT-UAM/CSIC-20-104}}}

\begin{abstract}
		
The recent electron recoil excess observed by XENON1T has a possible interpretation in terms of solar axions coupled to electrons. If such axions are still relativistic at recombination they would also leave a cosmic imprint in the form of an additional radiation component, parameterized by an effective neutrino number $\dnef$. We explore minimal scenarios with a detectable signal in future CMB surveys: axions coupled democratically to all fermions, axion-electron coupling generated radiatively, the DFSZ framework for the QCD axion. The predicted $\dnef$ is larger than $0.03-0.04$ for all cases, close to the $2\sigma$ forecasted sensitivity of CMB-S4 experiments. This opens the possibility of testing with cosmological observations the solar axion interpretation of the XENON1T excess.

\end{abstract}
		
	\maketitle
	
\section{Introduction}

Light and weakly coupled bosons appear naturally in several extensions of the Standard Model (SM). A famous example is the one of Peccei-Quinn (PQ) theories~\cite{Peccei:1977hh,Peccei:1977ur} with the QCD axion~\cite{Wilczek:1977pj,Weinberg:1977ma} as a pseudo-Nambu-Goldstone-boson (PNGB). More generally, PNGB are ubiquitous in motivated theoretical frameworks where axion-like particles (ALPs) arise from string compactifications~\cite{Arvanitaki:2009fg}. 

A promising strategy to discover these particles is by searching for electron recoils induced by the absorptions of axions produced in the Sun. Recently, the XENON1T experiment has reported an excess in the number of electron recoil events in the energy range $1 - 7 \, {\rm keV}$~\cite{Aprile:2020tmw}. Among several plausible explanations, solar axions stood up with a $3.5\sigma$ statistical significance. However, one should take the solar axion interpretation with the necessary caution. Other signal interpretations that do not require new physics, such as a higher concentration of tritium \cite{Aprile:2020tmw,Robinson:2020gfu,bhattacherjee2020xenon1t}, remain viable. Moreover, the value of the axion-electron coupling favored by XENON1T is in sharp tension with stellar cooling bounds~\cite{Aprile:2020tmw,DiLuzio:2017ogq} (See also Refs.~\cite{,Gao:2020wer,Sun:2020iim}), though some models appear to be able to escape them~\cite{bloch2020exploring}.

In this work, with the above caveats in mind, we correlate the solar axion interpretation of the XENON1T excess with a distinct cosmological signal. The observed events inform us that axions couple to electrons, and this leads to the natural expectation that it could couple to other SM fermions as well. We consider a few plausible examples where the axion: (i) couples to all SM fermions with the same strength; (ii) couples at tree level only to one SM fermion and this induces a nonzero coupling to electrons at one loop; (iii) is part of a well defined framework that indeed has couplings to all SM fermions, the DFSZ case~\cite{Dine:1981rt,Zhitnitsky:1980tq}. Such couplings with SM fermions may imply~\cite{Turner:1986tb,Ferreira:2018vjj,DEramo:2018vss} that axions achieve thermal equilibrium in the early universe, and later decouple from the primordial plasma at a temperature $T_D$. This gives rise to an additional radiation component, with observable effects on the CMB spectra, historically parameterized in terms of an effective neutrino number as
\begin{eqnarray}
\dnef = \left. 13.6 \, g_{*s}^{-4/3} \right|_{T=T_D} \ ,
\label{dnefgstar}
\end{eqnarray}
where $g_{* s}$ is the number of relativistic degrees of freedom contributing to the entropy density. This expression teaches us that the later the axion decouples, the larger the final $\dnef$ is. In many models the axion thermalizes well above the weak scale by interactions with gluons~\cite{Masso:2002np} or with the top quark~\cite{Salvio:2013iaa}, and for the above mentioned reason the final prediction will be dominated by the interactions with heavy fermions below ${\cal O}$(100) GeV. Such processes could potentially be probed with CMB-S4 experiments which have a forecasted sensitivity of $\sigma(\dnef) \simeq 0.027$ \cite{Abazajian:2016yjj} thus opening up a window to probe axion thermalization for temperatures below ${\cal O}$(100) GeV. In this paper we show how such production may be linked to the XENON1T excess in simple models.

We discuss axion production via fermion scattering in Sec.~\eqref{sec:Rates}, providing cross sections for the processes contributing to the signal in $\dnef$ and the related Boltzmann equations. We consider two main classes of explicit realizations: a non-anomalous ALP coupled to SM fermions in Sec.~\eqref{sec:ALP}, and the QCD axion in Sec.~\eqref{sec:QCD}. Measuring a non-vanishing contribution to $\dnef$ would provide the additional information discussed in Sec.~\eqref{sec:Consequences}. We defer radiatively induced axion couplings to App.~\eqref{app:RGE}, and we conclude in Sec.~\eqref{sec:Conclusion}.
	
\section{Thermal axions production via fermion scattering} 
\label{sec:Rates}

We consider axion production via fermion scattering below the electroweak phase transition (EWPT). These processes are mediated by the following dimension 5 contact interactions
\begin{eqnarray}
{\cal L}_{a\psi\psi} = \frac{\partial_\mu a}{2f} \sum_\psi c_\psi \bar{\psi} \gamma^\mu \gamma^5  \psi \ ,
\label{eq:La}
\end{eqnarray}  
with $a$ and $\psi$ the axion and SM fermions, respectively. The quantity $f$ is the axion decay constant, and we implicitly consider microscopic models where the only new degree of freedom accessible below the scale $f$ is the axion. The dimensionless coefficients $c_\psi$ encode unknown UV dynamics and can be thought as the result of integrating out heavy physics at energy scales above $f$. They are energy dependent and we provide details of their renormalization group evolution (RGE) in App.~\eqref{app:RGE}.

There are two leading production channels: fermion/antifermion annihilation ($\bar{\psi} \psi \; \rightarrow \; X a$) and Compton-like scattering ($\psi X \; \rightarrow \; \psi a$, and the same with the antifermion $\bar{\psi}$). The particle $X$ can be either a gluon or a photon depending on whether the SM fermion $\psi$ carries color charge. For colored fermions, namely SM quarks $q$, axion production is driven by processes with gluons, whose cross sections read~\cite{Ferreira:2018vjj}
\begin{eqnarray}
\label{eq:qqga} \sigma_{\bar{q} q \rightarrow g a}= & \, \dfrac{ c_q^2 g_s^2 x_q}{9 \pi f^2 } \dfrac{\tanh ^{-1}\left(\sqrt{1-4x_q}\right)}{1-4 x_q} \ , \\
\label{eq:qgqa}  \sigma_{q g \; \rightarrow \; q a}= & \, \dfrac{c_q^2 g_s^2 x_q}{192 \pi f^2 }  \dfrac{4x_q - 2 \ln \left(x_q\right)-x_q^2 -3}{1-x_q}.
\end{eqnarray}
Here, $x_\psi = m_\psi^2/E^2_\text{CM}$ and $E_\text{CM}$ is the energy in the center of mass frame. If the SM fermion responsible for axion production is a lepton, processes with photons would dominate and they have cross sections~\cite{DEramo:2018vss}
\begin{eqnarray}
\label{eq:llgammaa} \sigma_{\ell^+ \ell^- \rightarrow \gamma a}= & \, \dfrac{c_\ell^2 e^2 x_\ell}{4 \pi  f^2} \dfrac{\tanh ^{-1}\left(\sqrt{1-4 x_l}\right)}{1-4 x_l} \ , \\
\label{eq:lgammala} \sigma_{\ell^{\pm} \gamma \rightarrow \ell^{\pm} a}= & \,\dfrac{c_\ell^2 e^2 x_\ell}{32 \pi  f^2} \dfrac{4 x_\ell - 2 \ln(x_\ell) - x_\ell^2 - 3}{1 - x_\ell} .
\end{eqnarray}

As it is manifest from these expressions, cross sections are proportional to $m_\psi^2$ and therefore lighter fermions need a smaller $f / c_\psi$ to thermalize. Since $\dnef \propto g_{* s}(T_D)^{-4/3}$, the later the axion decouples the larger the signal would be; processes with light fermions, if efficient, would give the leading contribution.

We compute the resulting $\dnef$ by solving the Boltzmann equation for the comoving density $Y_a= n_a/s$, where $n_a$ is the axion number density and $s=2\pi^2 g_{*s}T^3/45$ the entropy density. We employ as a ``time variable'' the dimensionless combination $x = M / T$, where $M$ is some convenient mass scale (e.g., the mass $m_\psi$ of the fermion under consideration), and the axion comoving density evolves according to 
\be \label{Boltzmann eq.}
\frac{dY_a}{dx} = \left(1- \frac{1}{3} \frac{\partial \ln g_{*s}}{\partial \ln x}\right)  \frac{\sum_S \gamma_S}{s H x} \left(1- \frac{Y_a}{Y_a^\text{eq}}\right) \ .
\ee
Here, $Y_a^\text{eq}$ is the axion equilibrium comoving density and the collision rate $\gamma_S$ for a specific process results in $\gamma_{i j \leftrightarrow k a} \equiv n_i^\text{eq} n_j^\text{eq} \left<  \sigma_{i j \rightarrow k a} v_\text{rel} \right>$. At low enough temperatures, the axion comoving density freezes to a constant value $Y^\infty_a$ and the resulting value of $\dnef$ reads
\begin{eqnarray}
\dnef = 74.85 \,  \left(Y^\infty_a\right)^{4/3} \ .
\end{eqnarray}

We provide a perturbative description of binary collisions producing axions, and such a description is valid as long as SM gauge couplings are small and the thermal bath is described by a weakly-coupled plasma of quarks and gluons. This clearly breaks down as we approach the QCD phase transition, and we conservatively stop the evolution described by the Boltzmann equation at $1 \, {\rm GeV}$. For production driven by the top quark as well as for the one driven by leptons this is not an issue. However, for the bottom and for the charm this is a potential serious problem. In particular for the charm, axion production is likely to be efficient also below the GeV scale, and since $g_{* s}$ is rapidly changing around that temperature this translates into a significant theoretical uncertainty on the amount of axions. Production via pion scattering~\cite{Chang:1993gm,Hannestad:2005df,Archidiacono:2013cha,Millea:2020xxp} could give some additional contribution for the values of  $f$ we are interested in~\footnote{Note however that for $f \gtrsim 5\times 10^7$ GeV, which is the case for the parameter space region analysed in this paper, Eq.~\eqref{gae_Xenon1T}, the rates given in~\cite{Hannestad:2005df} would give decoupling temperatures above 200 MeV. For these temperatures we enter the QCD phase transition where we cannot assume the existence of thermal pions.}. For these reasons, we should interpret the output of our calculations for bottom and charm as a lower bound on the resulting $\dnef$. Moreover, we assume zero axion abundance at temperatures slightly above the EWPT. An initial abundance could also be present but that would depend on other aspects such the reheating temperature and on the value of $g_{* s}$ at higher temperatures~\footnote{Assuming an initial axion abundance due to scatterings that decouple at $T_D\gg {\cal O} (100)$ GeV~\cite{Masso:2002np,Salvio:2013iaa} would simply flatten the curves of the $\dnef$ predictions at large $f$ to the equilibrium value, which is at most the value obtained assuming the SM with no extra degrees of freedom at such $T_D$, $\dnef\simeq 0.027$, see~\cite{Ferreira:2018vjj,DEramo:2018vss}.}.

In the next two sections, we study these processes in various setups corresponding to different choices for UV fermion couplings as well as different relations between them and the ones to SM gauge bosons. 

 \section{Non-anomalous ALPs}
 \label{sec:ALP} 
  
We consider ALPs arising from the spontaneous breaking of a non-anomalous symmetry. For this reason, at the symmetry breaking scale $f$ we have only the couplings to fermions in Eq.~\eqref{eq:La} and no couplings to gauge bosons. Nevertheless, dimension 5 couplings to gauge bosons can be generated as a consequence of threshold corrections, proportional to $(m_a/f)^2$, once we integrate out SM fermions~\cite{Bauer:2017ris}. ALPs contributing to dark radiation must be relativistic between the epoch of matter-radiation equality and recombination. This results into the upper bound $m_a \lesssim {\cal O}(0.1)$eV, and threshold corrections are negligible for these masses. We account for coupling to photons in the next section when we study the QCD axion case.

\begin{figure}
	\includegraphics[width=1\linewidth]{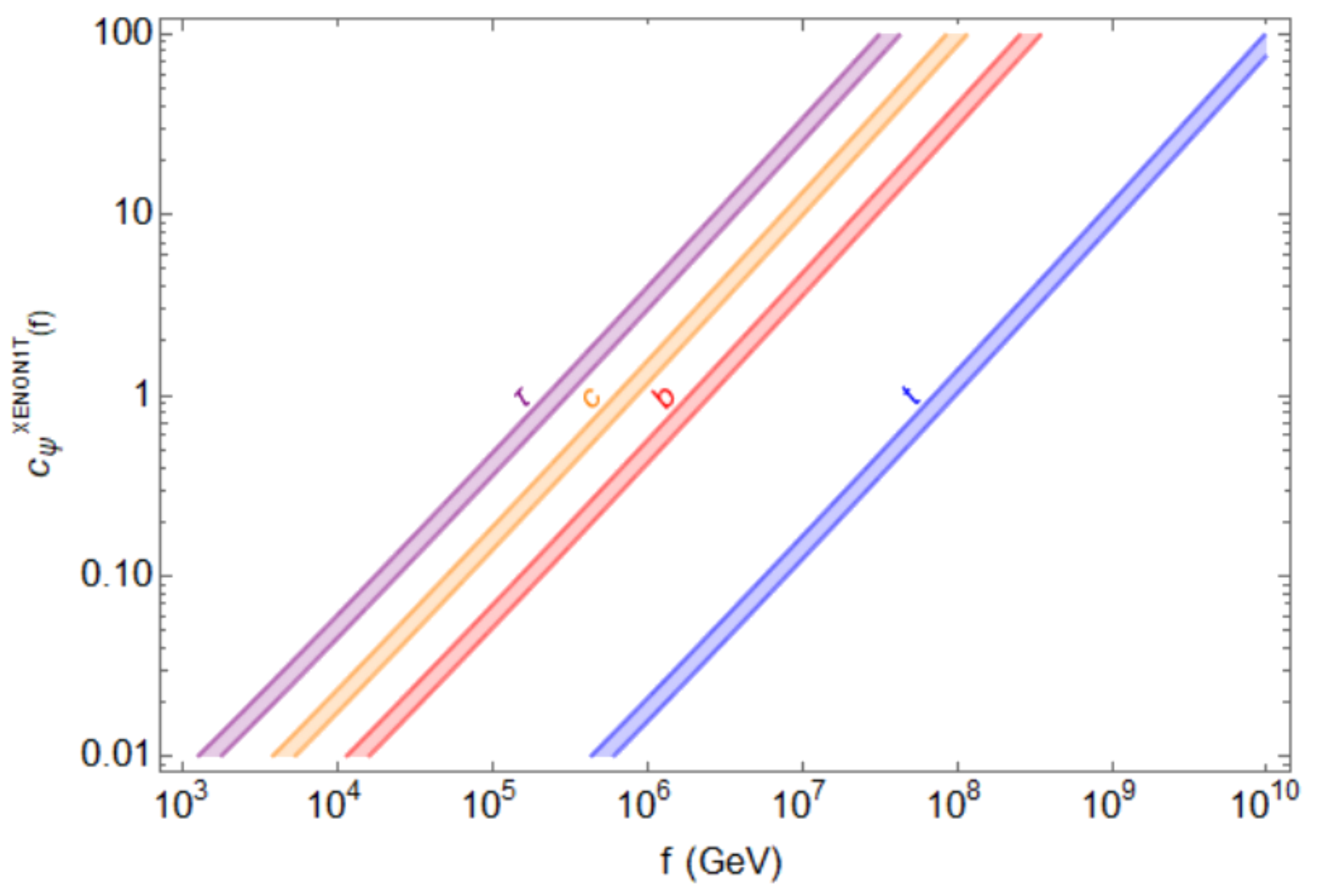}
	\caption{\it Relation between the axion-fermion coupling $c_\psi(f)$ at the UV scale and the scale $f$ itself that we need in order to generate a radiative coupling to the electron consistent with the XENON1T excess.}
	\label{fig:AxionRGE}
\end{figure}

We define each case studied in this section by a choice of Wilson coefficients $c_\psi(f)$ at the UV scale $f$. The resulting couplings at low energy, $c_\psi$, can be found according to the RGE prescription provided in App.~\eqref{app:RGE}. The low-energy axion-electron interaction that we need in order to address the XENON1T excess lies in the range
\be \label{gae_Xenon1T}
\left. \frac{f}{c_e} \equiv \frac{m_e}{g_{ae}}\right|_{\rm XENON1T}  \simeq (1.46 - 1.96) \times 10^8 \, {\rm GeV} \ .
\ee
This could turn out to be the case both because the axion couples to electrons at the high scale $f$ ($c_e(f)\neq 0$), or because the low-energy coupling $c_e$ is induced by radiative corrections. In the latter case, we need the fermion couplings at the UV scale illustrated in Fig.~\ref{fig:AxionRGE}. 

\subsection{Democratic ALP} \label{sec: Democratic ALP}
  
The first case we consider is the \textit{democratic} ALP where the axion has democratic ($c_\psi  \sim 1 $) and flavor conserving couplings to all fermions; constraints on flavor violating couplings~\cite{MartinCamalich:2020dfe} are too stringent and they do not allow a feasible explanation of the XENON1T excess. This scenario can be motivated in two different ways: one can assume that all fermions have couplings of order one in the UV, or one can consider an axion-top coupling of order unity in the UV ($c_t(f) \sim 1 $) and RGE would generate axion couplings $c_\psi\sim 1$ at low energy to all fermions.

\begin{figure}
	\centering
	\includegraphics[width=1\linewidth]{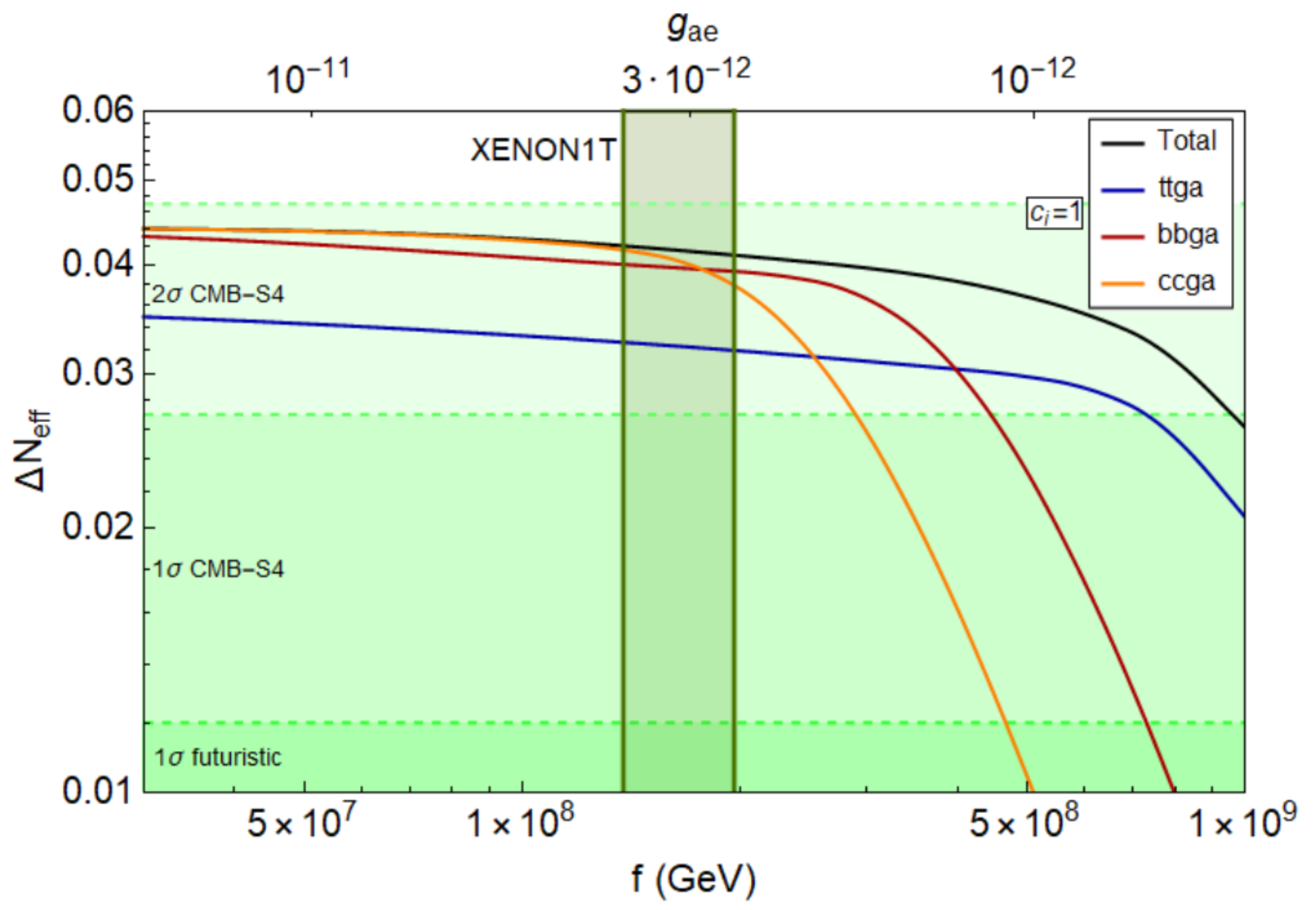}
	\caption{\it  $\dnef$ in the democratic case where the relevant channels are scatterings with heavy quarks $c,b,t$. We assumed no initial axion abundance above the EWPT and integrated the Boltzmann equation down to $1$~GeV to avoid getting too close to strongly coupled regimes. Green bands represent the forecasted sensitivity of CMB-S4  experiments~\cite{Abazajian:2016yjj}. Notice that the XENON1T window is in tension with the bound $f\gtrsim1.9\times 10^9 \GeV$ coming from stellar cooling~\cite{DiLuzio:2020jjp}.}
	\label{fig:neffquarkgluonscatterings}
\end{figure}

We set $c_\psi=1$ and the signal in $\dnef$ is dominated by axion-heavy quark scatterings. We solve the Boltzmann equation with the cross-sections given in Eqs.~\eqref{eq:qqga} and \eqref{eq:qgqa} assuming zero axion abundance at temperatures above the EWPT.  The results are shown in Fig.~\ref{fig:neffquarkgluonscatterings}. For $f$ in the XENON1T window the scatterings with the charm and bottom dominate the signal and yield $\dnef$ slightly above $0.04$. 
  
\subsection{Loop-induced electron coupling}

\begin{figure*}
	\centering
            \includegraphics[width=0.49\textwidth]{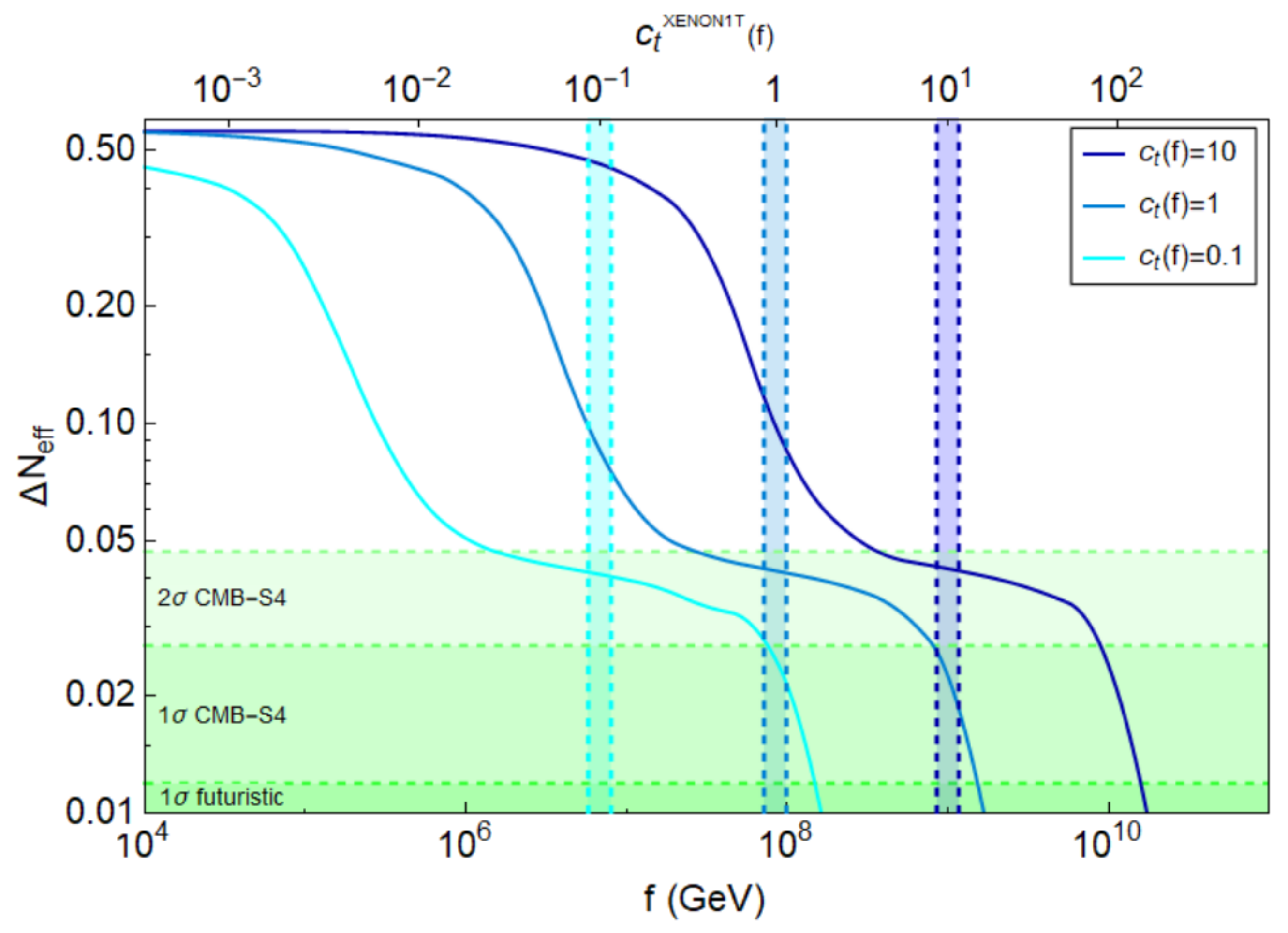} $\;$
            \includegraphics[width=0.49\textwidth]{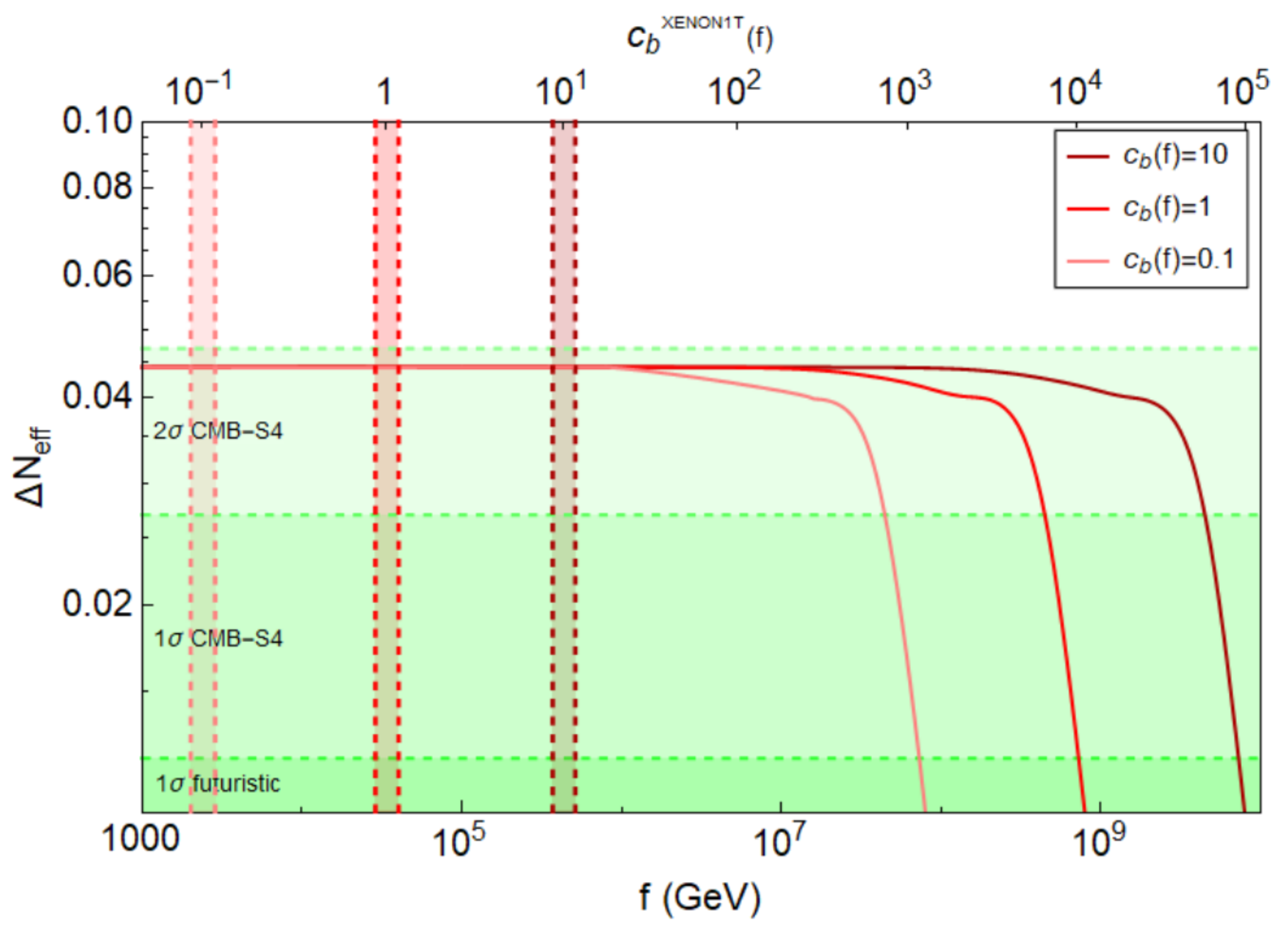}     
            \includegraphics[width=0.49\textwidth]{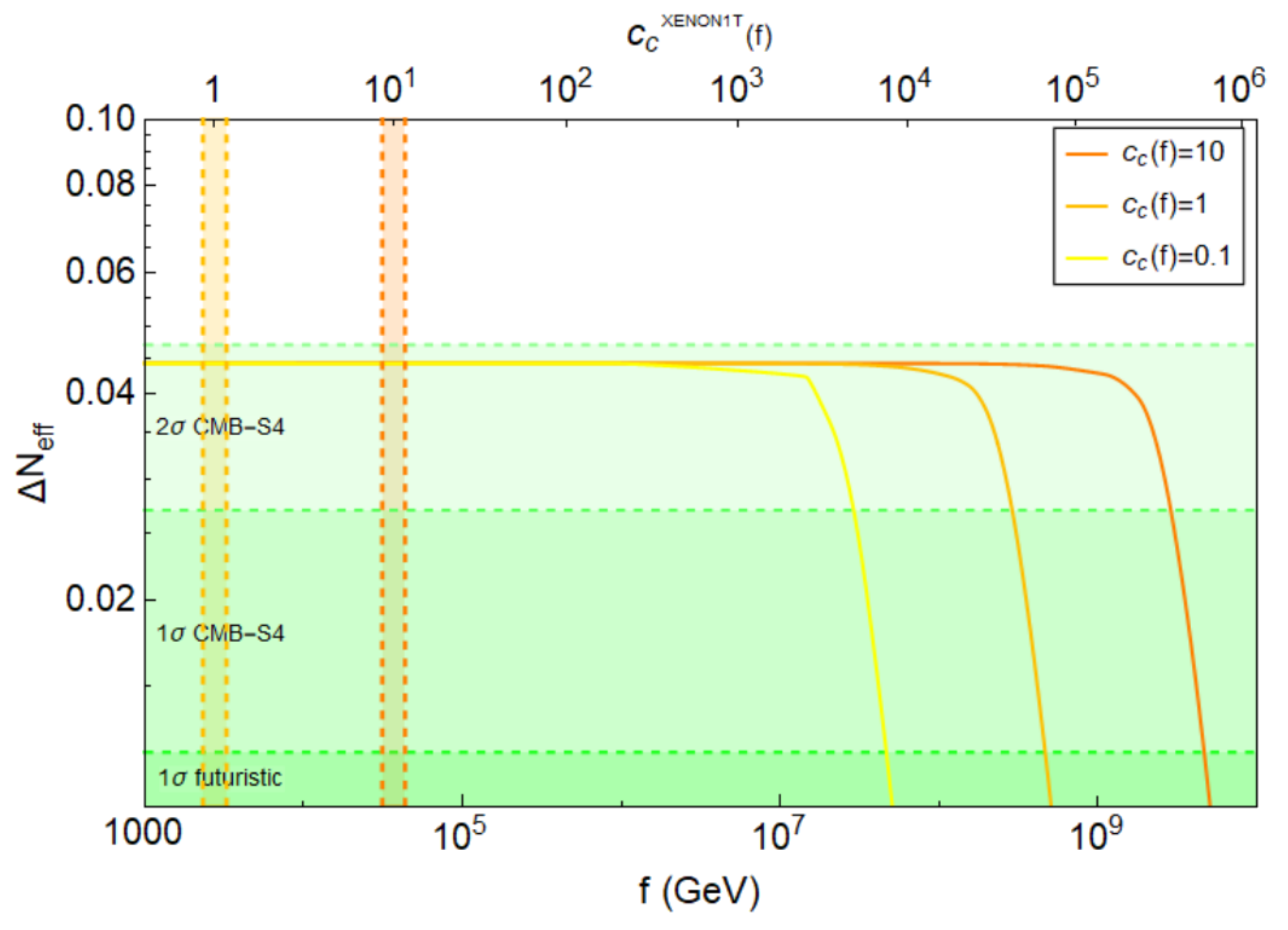}  $\;$
             \includegraphics[width=0.49\textwidth]{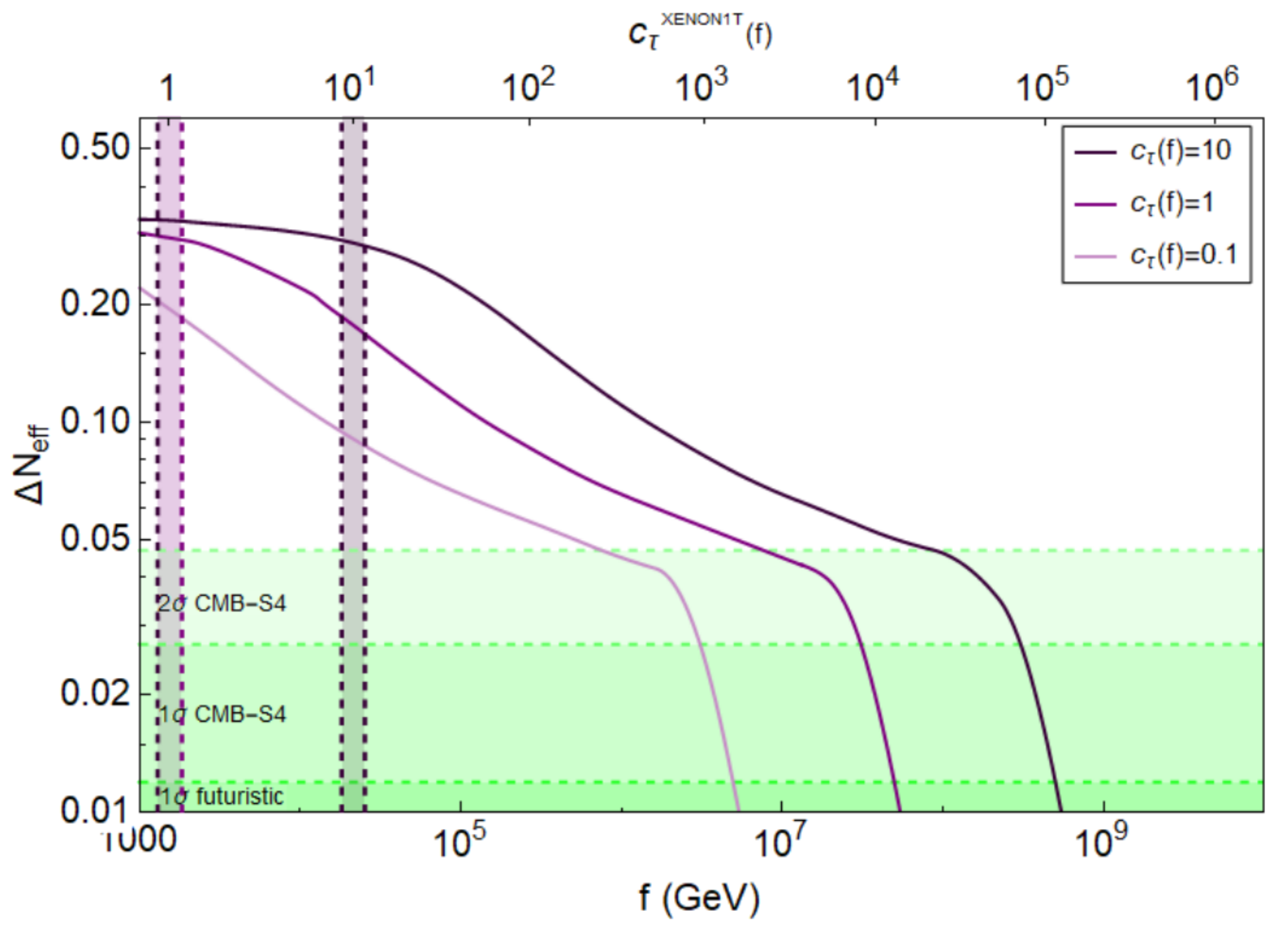}
	\caption{\it $\dnef$ as a function of $f$ for a few values of $c_\psi$(f) in the scenario where the axion-electron coupling is generated at loop-level. We assumed no initial abundance of axions above the EWPT. For quarks we stopped the Boltzmann equation at $1$GeV to avoid getting too close to strongly coupled regimes. The upper horizonal axis indicates the value of $c_\psi(f)$ needed for any given $f$ to explain the XENON1T excess. Green bands are the forecasted sensitivity of CMB-S4 experiments~\cite{Abazajian:2016yjj}.  Notice that the XENON1T window is in tension with the bound $f/c_e\gtrsim1.9\times 10^9 \GeV$ coming from stellar cooling~\cite{DiLuzio:2020jjp}.}
\label{fig:neffleptonsquarks}
\end{figure*}

The second scenario we study is the one where the axion-electron coupling at low energies is radiatively induced from an axion-fermion coupling  ($\psi= \tau, c,b$ or $t$) at the UV scale. The values of $c_\psi(f)$ needed to explain the XENON1T excess are given in Fig.~\ref{fig:AxionRGE} for each fermion~\footnote{We do not consider the muon because the coupling needed to generate the correct $g_{ae}$ is of order $c_\mu(f)/f\sim 10^{-4} {\rm GeV}^{-1}$, which is disfavored by about a few orders of magnitude  by supernova constraints~\cite{Brust:2013ova,DEramo:2018vss,Bollig:2020xdr,Croon:2020lrf}.}. For each case, we assume a single $c_\psi(f)$ to be nonzero at the UV scale and we solve the Boltzmann equation including all the radiatively induced couplings at low energy. 

Our predictions for $\dnef$ as a function of $f$ for different values of $c_\psi(f)$ are shown in Fig.~\ref{fig:neffleptonsquarks}. In the upper horizontal axis we show the value of $c_\psi(f)$ needed to explain the XENON1T excess for the associated value of $f$. The predicted $\dnef$ for each fermion is quite sharp in this loop-induced scenario because $c_\psi(f)/f$ is mostly fixed by the RGE (up to logarithmic corrections and the experimental uncertainty in $g_{ae}$). 

In the case of the top, upper left plot, the XENON1T region corresponds to $\dnef \sim 0.04$ and $f = [6 \times 10^6, 10^9]$ GeV for $c_t(f)$ in the window $0.1-10$. Note that in this case the radiatively induced couplings to others fermions ($\mu, \tau, c$ and $b$) are relevant and so we accounted for several channels to produce the axion. 

For bottom and charm, respectively upper right and lower left plot, the axion is in thermal equilibrium at $1$ GeV for couplings in the XENON1T region. This is the temperature at which we stop the Boltzmann equation for quarks, and the relic abundance saturates at $\dnef=13.8\, g_{*s}^{-4/3}|_{T=1\text{GeV}} $ for lower values of $f$. Thus the prediction of  $\dnef \simeq 0.044$ should be understood for these cases as a lower bound on the signal. The range of $f$ needed to get the right loop-induced $c_e$ is, for the bottom, $f= \left[2 \times 10^3, 5 \times 10^5\right]$ GeV for $c_b(f) =[0.1,10]$ and, for the charm, $f= \left[ 2 \times 10^3, 4 \times 10^4\right]$ GeV in the window $c_c(f) =\left[1,10\right]$. 

Finally, we look at the $\tau$, lower right plot. This case is quite interesting because the axion thermalizes at a much lower temperature and the calculation is still under control since it does not involve QCD. The relative signal is boosted to values of $\dnef\simeq 0.3$ and $f= \left[  10^3, 3 \times 10^4\right]$ GeV for $c_\tau(f) =[1,10]$. 
Such a large value of $\dnef$ is already now within the $2\sigma$ sensitivity region of the latest CMB experiments. In particular, although CMB and LSS data alone do not hint at a non-zero value of $\dnef$ \cite{Aghanim:2018eyx}, when including SH$_0$ES 2019 local Hubble constant measurement of $H_0$~\cite{Riess:2019cxk} there is a shift of the central value towards $\dnef=0.26^{+0.16}_{-0.15}$~\cite{Ballesteros:2020sik} (or $\dnef=0.28^{+0.16}_{-0.17}$~\cite{Gonzalez:2020fdy} adding also the Pantheon Supernova dataset) which is in remarkable agreement with the above prediction.
Such values will be tested also by forthcoming CMB experiments, such as LiteBIRD~\cite{LITEBIRD},  Simons Observatory~\cite{Ade:2018sbj} and CMB-S4~\cite{Abazajian:2016yjj}.

\section{QCD axion} \label{sec:QCD}
	
In the QCD axion case the non-perturbative axion potential leads to the general relation for its mass~\cite{Bardeen:1978nq,diCortona:2015ldu}
\be
m_a=5.70(6)\times 10^{-2} \left(\dfrac{10^{8}\GeV}{f}\right) \text{eV} \,. \label{massequation}
\ee
For $f\sim 10^8$GeV the axion is relativistic at the time of CMB decoupling and thus will again contribute to $\dnef$.

There are two benchmark classes of QCD axion models: KSVZ~\cite{Shifman:1979if} and DFSZ~\cite{Dine:1981rt,Zhitnitsky:1980tq}. The former does not have tree-level couplings to SM fermions so it does not seem able to explain the XENON1T excess and satisfy the CAST bound at the same time \cite{Arik:2008mq,Aprile:2020tmw}. Therefore we focus on the DFSZ models whose couplings to  quarks satisfy to
\be
c_U+c_D=\dfrac13\,,
\ee
where $c_U$ is the universal coupling to the up-type quarks and $c_D$ the universal coupling to the down-type ones. The axion may couple to charged leptons as to the up-type quarks or as to the down-type quarks: we take the second option for concreteness, as in~\cite{Aprile:2020tmw}, i.e. $c_E=c_D$, being $c_E$ the universal coupling to the charged leptons. 

The DFSZ model also features two Higgs doublets, but the extra Higgs and also the rest of the SM couplings (i.e.~with gauge bosons and the physical Higgs) are neglected here, since they would affect axion production only at very high T and would give a subdominant contribution due to eq.~(\eqref{dnefgstar}), compared to the production via fermions, which are relevant at lower temperatures, $T_D\approx 1-10$ GeV. The photon-axion coupling in DFSZ model takes the value $c_{a\gamma\gamma}=8/3~(c_{a\gamma\gamma}=2/3)$ if the charged leptons couple to the axion as the down-type (up-type) quarks do; such a coupling is important for experiments that search for axions, but it  gives a subdominant contribution to $\dnef$. Finally, there are no RGE effects in this case since we can always choose a basis where the axion appears only inside the quark mass matrix~ \cite{diCortona:2015ldu}.

\begin{figure}
	\centering
	\includegraphics[width=1\linewidth]{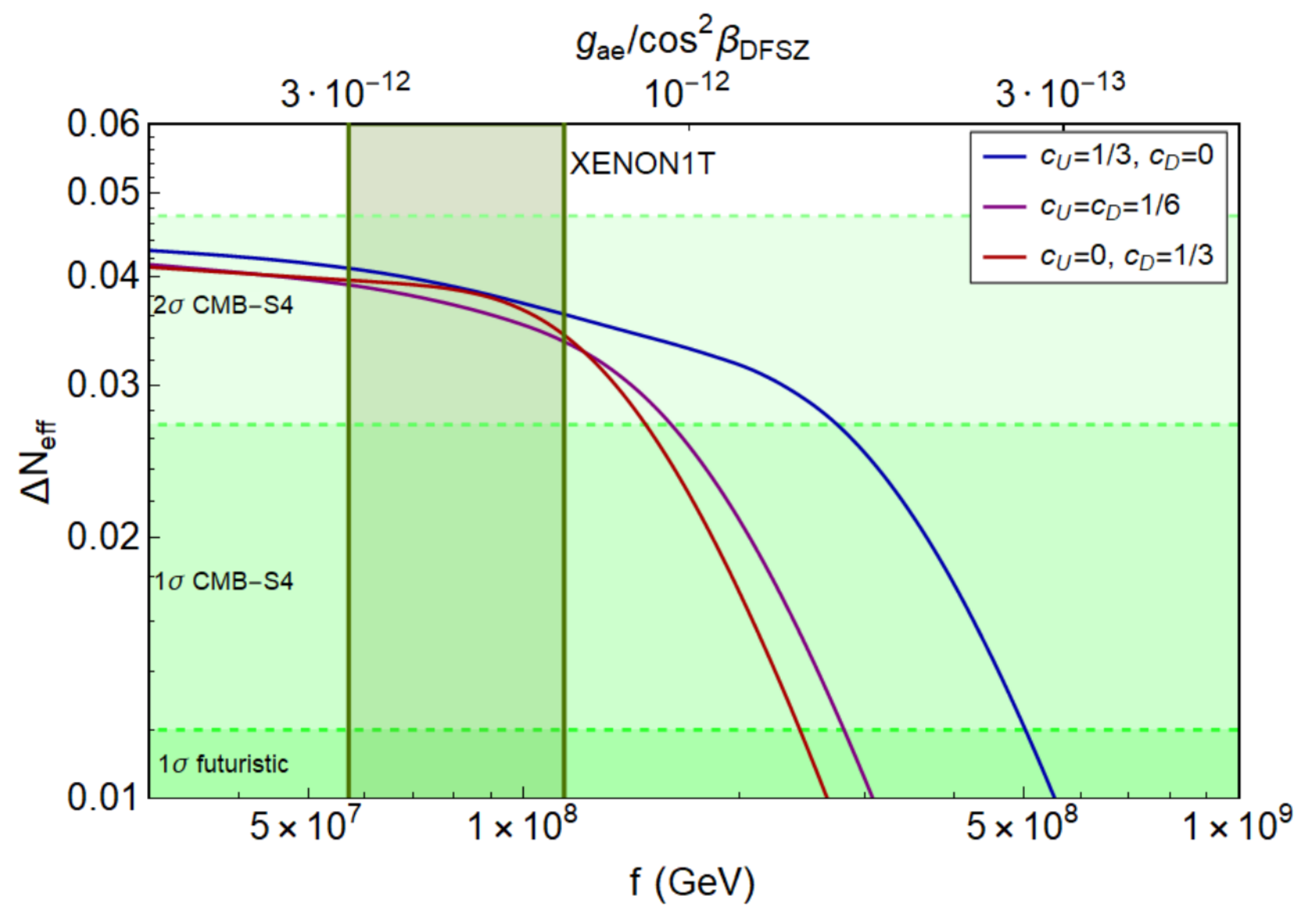}
	\caption{\it Prediction for $\dnef$ for the DFSZ axion model. Here we show the top, bottom and charm contributions for three different choices of the PQ charges: $c_U = 1/3$ and $c_D =0$ in blue, $c_U =0$ and $c_D =1/3$ in red, and $c_U =1/6=c_D$ in purple. We assumed no initial abundance of axions at low temperatures and integrated the Boltzmann equation down to $1$GeV to avoid getting too close to strongly coupled regimes. Green bands are the forecasted sensitivity of CMB-S4 experiments~\cite{Abazajian:2016yjj}.  The XENON1T window is in tension with the bound $f/c_E\gtrsim1.9\times 10^9 \GeV$ coming from stellar cooling~\cite{DiLuzio:2020jjp}, but escapes the limit $f\gtrsim 1.2\times 10^7\GeV$ set by CAST~\cite{Arik:2008mq}. The XENON1T window for this model could however be just perhaps reached by BabyIAXO (reaching $f\sim 5.7\times 10^7 \GeV$) and fully explored by the IAXO experiment~\cite{Armengaud:2019uso}, that could reach $f\sim 2\times 10^8 \GeV$.}
	\label{fig:neffxenondfsz}
\end{figure}

The contributions to $\dnef$ are dominated by scatterings with heavy quarks and can be seen in Fig.~\ref{fig:neffxenondfsz}, where three cases are considered: $c_U = 1/3$ and $c_D =0$ in blue in the plot, $c_U =0$ and $c_D =1/3$ in red, and $c_U =1/6=c_D$ in black. While the vertical axis represents $\dnef$, the horizontal one stands for $f$ or equivalently for $g_{ae}/\cos^2\beta_\text{DFSZ}$, being $\cos^2\beta_\text{DFSZ}=x^2/(x^2+1)$ the parameterisation of $x=v_1/v_2$, which is the ratio of the Higgs VEVs.

\boldmath
\section{What $\dnef>0$ can teach us} \label{sec:Consequences}
\unboldmath

Having correlated the signal in XENON1T with a potential non-vanishing $\dnef$, it is now instructive to ask what we can learn from a non-zero detection of $\dnef$:

\paragraph{No detection, $\dnef \lesssim 0.03$:}
The axion does not couple to heavy quarks or the coupling is small.  

\paragraph{$\dnef \sim 0.03-0.05$:}
A detection of $\dnef$ in this window would give a strong hint that the axion couples to at least one of the heavy quarks.
In particular, assuming that we know $g_{* s}(T_D)$ with enough precision from latest lattice simulations \cite{Borsanyi:2016ksw} then in a given model where all the PQ charges are fixed, e.g. DFSZ, the detection in XENON1T would tell us the value of $f$ and, consequently, there would be a sharp prediction for $\dnef$. Therefore, in principle one would be able to test if the ratio of PQ charges $c_\psi/c_e$ is indeed the predicted one.

\paragraph{$\dnef \gtrsim 0.05$:}
In this case axion production could come from different sources: either from the $\tau$ with a small $f$ or from axion coupling to charm or bottom at temperatures below $1$ GeV. In the latter case a reliable calculation of axion production close to the QCD phase transition would be needed.

	\section{Conclusions} \label{sec:Conclusion}

Next-generation detectors such as XENONnT and others \cite{Zhang:2018xdp,Akerib:2019fml} will be able to discriminate with high significance between the different interpretations for the excess in the number of electron recoil events at XENON1T. The solar axion interpretation still has to overcome the challenge of being compatible with stellar cooling results. However, if it remains firm, it will open a whole new axion window to the universe.

In this work we explored correlated signals of the XENON1T excess in cosmological data. Namely, we studied how the presence of axion couplings to other leptons and quarks can give a sizable contribution to $\dnef$ due to the possibility of thermalizing the axion at temperatures of order $1-10$ GeV. We presented three different motivated setups where such couplings would exist: i) if the axion couples democratically to all fermions in the UV (Fig.~\ref{fig:neffquarkgluonscatterings}); ii) if the axion-electron coupling compatible with XENON1T is radiatively induced from an axion-($\tau,c,b,t$) coupling (Fig.~\ref{fig:neffleptonsquarks}); iii)  the DFSZ model of the QCD axion (Fig.~\ref{fig:neffxenondfsz}). The largest signal comes from case ii) when the axion couples only to the $\tau$ in the UV. In such a case we find in the XENON1T region  $f/c_\tau \sim 10^3$ GeV and $\dnef \simeq 0.3$, which interestingly coincides with the recent CMB analyses including supernova data \cite{Ballesteros:2020sik,Gonzalez:2020fdy}. In the remaining cases the values of $f/c_\psi$ in the XENON1T region are considerably higher, up to $f/c_\psi \sim 10^8$ GeV, and the signal is predicted to be $\dnef \gtrsim 0.04$, which can still be detected at the $2\sigma$ level with future CMB-S4 experiments \cite{Abazajian:2016yjj}.
In all such cases the XENON1T range leads to the possibility of testing axion physics through cosmological data in the coming years, and would allow us to experimentally test the Universe at temperatures of $T\approx 1-10$ GeV. This would be a remarkable improvement over our current ability to look at the earliest stage of the universe, going by 3 or 4 orders of magnitude above the present highest experimentally accessible temperature, $T\approx $ MeV, given by nucleosynthesis.

	${}$\linebreak
	\emph{\textbf{Acknowledgments:}}
The work of A.N. work is supported by the grants FPA2016-76005-C2-2-P, PID2019-108122GB-C32, MDM-2014-0369 of ICCUB (Unidad
de Excelencia Maria de Maeztu), AGAUR2017-SGR-754.

F.A.A. and L.M. acknowledge partial financial support by the Spanish MINECO through the Centro de excelencia Severo Ochoa Program under grant SEV-2016-0597, from European Union's Horizon 2020 research and innovation programme under the Marie Sklodowska-Curie grant agreements 690575 (RISE InvisiblesPlus) and 674896 (ITN ELUSIVES), by the Spanish ``Agencia Estatal de Investigac\'ion''(AEI) and the EU ``Fondo Europeo de Desarrollo Regional'' (FEDER) through the projects FPA2016-78645-P and PID2019-108892RB-I00/AEI/10.13039/501100011033. L.M. acknowledges partial financial support by the Spanish MINECO through the ``Ram\'on y Cajal'' programme (RYC-2015-17173).

The work of F.D. is supported by the research grant "The Dark Universe: A Synergic Multimessenger Approach" number 2017X7X85K under the program PRIN 2017 funded by the Ministero dell'Istruzione, Universit\`a e della Ricerca (MIUR), by the University of Padua through the ``New Theoretical Tools to Look at the Invisible Universe'' project and by Istituto Nazionale di Fisica Nucleare (INFN) through the Theoretical Astroparticle Physics (TAsP) project.
\vspace{0.1cm}		

		\appendix
	
\section{RGE of axion couplings}
\label{app:RGE}

As already emphasized in Sec.~\eqref{sec:Rates}, the effective axion couplings to SM fermions in Eq.~\eqref{eq:La} are originated at the PQ breaking scale $f$ and at lower energies there are only SM fields and the axion itself. This working assumption allows us to evolve the dimensionless Wilson coefficients $c_\psi$ at lower energy scales by only using known SM interactions. We neglect flavor violating effects and neutrino masses, and the detailed RGE can be found in Refs.~\cite{Crivellin:2014qxa,DEramo:2014nmf} where anomalous dimension matrices are derived both above and below the EWPT. We do not consider the RGE of SM couplings and we limit ourselves to the results of a fixed-order calculation. The expressions for the low-energy couplings can be written in terms of simple analytical expressions~\cite{Feng:1997tn,DEramo:2016gos,DEramo:2017zqw}, and at a generic renormalization scale $\mu < f$ they read
\be
\begin{split}
c_\psi(\mu) = & \,  c_\psi(f) - T^{(3)}_\psi \, \times \, \\ & 
 \left[\sum_{m_{\psi^\prime} < \mu} c_\psi^\prime(f) \frac{N_{\psi^\prime}^{(c)} T^{(3)}_{\psi^\prime} \lambda_{\psi^\prime}^2}{2 \pi^2} \,\ln\left( \frac{f}{\mu} \right) + \right. \\ &
\left. \sum_{m_{\psi^\prime} > \mu} c_\psi^\prime(f) \frac{N_{\psi^\prime}^{(c)} T^{(3)}_{\psi^\prime} \lambda_{\psi^\prime}^2}{2 \pi^2} \,\ln\left( \frac{f}{m_{\psi^\prime}} \right) \right] \ .
\end{split} 
\ee
Here, $T^{(3)}_\psi = (+1/2 , - 1/2)$ is the value of the third component of the weak-isospin for the fermion $\psi$, that also has a Yukawa coupling $\lambda_\psi$ and number of colors $N_{\psi}^{(c)}$. The first sum runs over SM fermions with mass below the renormalization scale $\mu$, and thus over degrees of freedom still accessible, whereas the second sum runs over SM fermions heavier than $\mu$ that have been integrated out.

In order to address the XENON1T excess, we need a significant low-energy coupling to electrons and its explicit expression in terms of UV couplings results in
\be \label{running of c_e}
c_e = c_e(f) + \sum_{\psi^\prime} c_\psi^\prime(f) \frac{N_{\psi^\prime}^{(c)} T^{(3)}_{\psi^\prime} \lambda_{\psi^\prime}^2}{4 \pi^2} \,\ln\left( \frac{f}{m_{\psi^\prime}} \right)\ ,
\ee
with the sum over all SM fermions coupled to the axion.

\bibliographystyle{BiblioStyle}
\bibliography{Xenon-Neff.bib}
	
\end{document}